\documentstyle[12pt]{article}
\begin{document}
\begin{titlepage}
\begin{center}
{\Large\bf Turbulence as a constrained system}
\vskip.1in 
A. C. R. Mendes\footnote{Email: \tt albert@fisica.ufjf.br}$^a$, W. Oliveira\footnote{Email: \tt oliveira@fisica.ufjf.br}$^b$ and F.I. Takakura\footnote{Email: \tt takakura@fisica.ufjf.br}$^b$\\~\\ 
$^a$ Centro Brasileiro  de Pesquisas F\'\i sicas, Rio de Janeiro, 
22290-180, RJ, Brasil,fax:+55+32 229-3312, phone:+55+32 229-3307 \\

$^b$ Departamento de F\'\i sica, ICE, Universidade Federal de Juiz de Fora,\\
36036-330, Juiz de Fora, MG, Brasil
\end{center}

\begin{abstract}
Hydrodynamic turbulence is studied as a constrained system from the point of view of metafluid dynamics. We present a Lagrangian description for this new theory of turbulence inspired from the analogy with electromagnetism. Consequently it is a gauge theory. This new approach to the study of turbulence tends to renew the optimism to solve the difficult problem of turbulence. As a constrained system, turbulence is studied in the Dirac and Faddeev-Jackiw formalisms giving the Dirac brackets. An important result is that we show these brackets are the same in and out of the inertial range, giving the way to quantize turbulence.
\end{abstract}
\leftline{PACS numbers: 47.27.-i, 11.90.+t}
\leftline{Keywords: Constrained Systems, Gauge Theory, Turbulence}
\end{titlepage}

%\pacs{PACS numbers: 47.27.-i, 11.90.+t} 

%\newpage 
\section{Introduction} 

Since its discovery, turbulence has attracted an enormous effort from scientists to study it \cite{Landau,Monin-Yagolm,Frisch}. In recent years, there has been an increasing interest in studying turbulent hydrodynamics because lots of phenomena in Nature are turbulent, like in the flow of blood in biomechanics, meteorology and ocean engineering, astrophysics and formation of galaxies in the Universe. Beyond these phenomena, turbulence has became a very fruitful field of research for theoreticians, who have studied the analogies between turbulence and field theory, critical phenomena and condensed matter physics \cite{Bramwell-Holdsworth-Pinton,L'vov,Periwal,Polyakov,Gurarie,Eyink-Goldenfeld,Nelkin,deGennes,Rose-Sulem} leading to renewed optimism to solve the problem of turbulence described by the Navier-Stokes equations \cite{Landau}
\begin{equation}
{\partial \vec u \over{\partial t}} = - \vec w \times \vec u - \nabla \left({p\over \rho} + {u^2\over 2}\right) + \nu \nabla^2 \vec u, \label{NS}
\end{equation}
where $\vec u (\vec  x, t)\;$ is the velocity field, $\vec w (\vec x, t)\;$ the vorticity field, $p(\vec x, t)\;$ is the pressure, $\rho\;$ is the  density and $\nu\,$ the kinematic viscosity.

Among turbulent phenomena, the hydrodynamic turbulence is one in which physicists have enormous interest because of the universal characteristics stressed by an incompressible fluid at high Reynolds numbers in the fully-developed turbulent regime. The regime related to the limit of high Reynolds number, $R \,\rightarrow\, \infty$, with $R \equiv (LV)/\nu$, which measures the competition between convective and diffusive processes in an incompressible fluid described by the Navier-Stokes equations. Above, $L\;$ is the integral length-scale of the largest eddies that should appear and $V$, is a characteristic large-scale velocity. 

An important feature in turbulence is the presence of vorticities \cite{Moriconi-Takakura}, but another flow quantity, the Lamb vector, can be used to describe turbulence\cite{Wu}, together with vorticities \cite{Marmanis} in the so called {\it metafluid dynamics}. In metafluid dynamics, the study of average quantities of an incompressible fluid in the fully developed regime is proposed, given a system where the average fields show up in a continuum  inter-relation and respond as waves to quantities named turbulent sources. To do so, the Lamb vector $\vec l\,$ and the vorticity become the kernel of the turbulent dynamics, being related by $\vec l(\vec x, t) = \vec w\times \vec u$. Defining Bernoulli energy function $\phi$, $\phi(\vec x, t) = {p\over \rho} + {u^2\over 2}$, one can introduce a theoretical concept, the turbulent charge density $n$, $\,n(\vec x, t) = - \nabla^2\phi$. In the inertial range, the Bernoulli energy function is conserved, the energy generated at larger scales is transferred to smaller wave numbers across the region and energy pumping and dissipation are not relevant allowing us to make a very close analogy between turbulent hydrodynamics and electromagnetism \cite{Marmanis}. Later on, we show there is no loss of generality in treating the turbulent flow in inertial range from the point of view of our proposal. 

This analogy between  hydrodynamic turbulence and the electromagnetism in inertial range makes possible to study metafluid dynamics. So as in the electromagnetism \cite{Jackson}, where we write the Lagrangian density as ${\cal L} = {1\over 2} (\vec E^2 - \vec B^2)\,$ ($\vec E\;$ is the electric field and $\vec B\;$ is the magnetic field), one can write down the Lagrangian density for the theory of turbulence as ${\cal L} = {1\over 2} (\vec l^2 - \vec \omega^2)\,$ or ${\cal L} = - {1\over 4}F_{\mu\nu} F^{\mu \nu} - J_\mu V^\mu $, with $F^{\mu \nu} = \partial^\mu V^\nu - \partial^\nu V^\mu, \;
V^\mu = (\phi, \,\vec u ), \hbox{ and }
J^\mu = (n, \, \vec J)$, $\vec J = \vec u n + \nabla \times (\vec u . \vec \omega) \vec u + \vec \omega \times \nabla (\phi + \vec u^2 ) + 2 (\vec l . \nabla) \vec u\;$ is the turbulent current density. Writting $\vec J = \vec J_{tr} + \vec J_l$, where $\vec J_{tr}\,$ is the turbulent transverse current density and $\vec J_l = \nabla \dot \phi\,$ is the turbulent longitudinal current density, the Lagrangian density becomes ${\cal L} = - {1\over 4} F_{\mu \nu} F^{\mu\nu} - J_\mu V^\mu + \vec u . \nabla \dot \phi$, where now $J^\mu = (n, J_{tr}^i)$. We note that $\nabla . \vec J_{tr}= 0 $.

Despite of all resemblance between hydrodynamic turbulence and the dynamics of electromagnetic fields, there is a conceptual difference in the identification of the physical entities. In the classical electromagnetism, the physical fields are the electric and magnetic fields and the potentials are just mathematical artifices while in the metafluid dynamics the potentials play physical role.

The analogy between turbulent hydrodynamics and electromagnetism suggests we can study the turbulence as a constrained system. The theory of constrained systems plays an important role in the study of physical systems. It includes almost all known fundamental theories. Constrained systems were first studied systematically by Dirac \cite{Dirac}.

Constrained systems are characterised in phase space by the presence of constraints, which are functions of the coordinates and momenta. These constraints can be classified into primary, secondary, etc., or first and second class \cite{Dirac,Teitelboim,Sundermeyer}. First class constraints imply the presence of gauge invariance in the theory, since they generate gauge transformations. Beside that, they exhibit the structure of the corresponding gauge group. On the other hand, second class constraints do not have these properties.

There exist several methods to treat constrained systems based on the classification given above. Most of these deal with first class constraints, for the second class ones there is the Dirac bracket method. There is also a more recent method by Faddeev and Jackiw \cite{Faddeev-Jackiw}, which does not follow the above classification. In this method, the Dirac formalism can be avoided and basic bracket relations can be obtained without using the usual Dirac brackets.

We will consider the Dirac and Faddeev-Jackiw formalisms to treat turbulence as constrained system. Those procedures have been used with great success in Quantum Field Theory for quantizing some models\cite{Barcelos-Wotzasek,Wilson}.

\section{Dirac formalism}

Under the point of view of a geometric interpretation, Arnold \cite{Arnold} showed, using Lie algebra, that Euler flow can be described in the Hamiltonian formalism in any dimension. This has a lot of interesting consequencies for fluid mechanics and has been studied \cite{Arnold-Khesin,Zeitlin}, but it is not quite obvious that this method could be used for viscous fluid. That is where the metafluid dynamics comes, showing a way to find a Hamiltonian formalism  even for turbulent flow with viscosity. 

Using again the analogy between electromagnetism and turbulence, one obtains the conjugate momentum for ${\cal L}\;$ as $\pi^\mu = {\partial {\cal L} \over{\partial \dot V_\mu}}= F^{\mu 0}\;$ that immediately leads us to a primary constraint $\pi^0 \approx 0 $. So, the primary Hamiltonian becomes $
H_p = \int {\rm d} \vec x \left( \pi^\mu \dot V_\mu - {\cal L} + \lambda \pi^0 \right)$ $= \int {\rm d} \vec x \left({1\over 2} \vec \pi^2 + {1\over 2}(\nabla \times \vec u)^2 + 2\vec \pi . \nabla \phi \right. $ $\left.+ \pi^0 \dot\phi + {1\over 2} (\nabla \phi)^2 + J^0 \phi - \vec J_{tr}. \vec u + \lambda \pi^0\right)$.

In order to apply the Dirac Hamiltonian method \cite{Dirac}, we need to look for secondary constraints. Imposing that primary constraint must be conserved in time, we get $\dot \pi^0 = \{ \pi^0, H_p\} \approx 0$. This consistency condition over the primary constraint leads to the secondary constraint $\nabla . \vec \pi - J^0 \approx 0 $, that is exactly one of the metafluid dynamics equations. If one goes on and imposes the consistency condition over this secondary constraint, we observe that no new constraints are obtained via this iterative procedure. 

The constraints $\pi^0 \approx 0\,$ and $\nabla . \vec \pi - J^0 \approx 0\,$ are first class constraints. It means this new theory of turbulence is gauge-invariant. A result expected due to the analogy between electromagnetism and turbulence. This analogy allows us to apply to this new theory of turbulence all the machinery associated with gauge theories. Since we have two constraints of first class, we have two degrees of freedom to field $V^\mu$. These two degrees of freedom can be associated to the vorticity. This means we must explore the gauge invariance of the theory to choose two gauges. Following the analogy between electromagnetism and turbulence we find the gauges $\nabla . \vec u = 0, \quad \hbox{(Coulomb or transverse gauge) and } V^0 = \alpha\hbox{ (constant)} \quad \hbox{(Temporal gauge)}$, where the first gauge comes from condition of incompressibility of fluid and the second, from the fact that Bernoulli energy function is constant.

So, the theory has the constraints $T_1 = V^0 - \alpha \approx 0,\;
T_2 = \pi^0 \approx 0, \;
T_3 = \nabla . \vec u \approx 0, \hbox{ and }
T_4 = \nabla . \vec \pi - J^0 \approx 0$,which are second class ones. To implement the Dirac brackets we need to compute the matrix elements of their Poisson brackets, which reads $C_{12} = \delta(\vec x - \vec y) = -C_{21},\;
C_{34} = - \nabla^2 \delta (\vec x - \vec y) = - C_{43}$, and all other elements are zero. Using the Dirac brackets $\{A, \, B\}^* = \{A, \, B\} - \{A, \, T_\alpha \}C_{\alpha \beta}^{-1} \{T_\beta, \, B\}$, we get
\begin{eqnarray}
\{u_i(\vec x), u_j(\vec y)\}^* &=& 0 = \{ \pi_i(\vec x), \, \pi_j(\vec y)\}^*, \nonumber \\
\{u_i(\vec x), \pi_j(\vec y)\}^* &=& \left(\delta_{ij} - {\partial_i^x \partial_j^x\over{\nabla^2}}\right) \delta(\vec x - \vec y).\label{i17}
\end{eqnarray}

Inspite of these Dirac brackets have been computed in the inertial range, they are valid  for all fully-developed turbulence, as we show bellow.

\section{Faddeev-Jackiw formalism}

With the purpose to show that the presence of viscous term in the Navier-Stokes equation does not change the Dirac brackets, let's analyze the metafluid dynamics using a second treatment for this constrained theory, the Faddeev-Jackiw procedure, also known as sympletic formalism. 

Faddeev and Jackiw \cite{Faddeev-Jackiw} showed how to implement the constraints directly into the canonical part of the first order Lagrangian.  Once this method is applied to first order Lagrangians, for implementing the Faddeev-Jackiw  procedure we write the Lagrangian density of turbulence making explicit the time-derivative of the fields $V_\mu$ :
\begin{eqnarray}
{\cal L} &=& {1\over 2} \dot{\vec u} + \dot{\vec u}. \nabla \phi + {1\over 2}(\nabla \phi)^2 - {1\over 2}(\nabla \times \vec u)^2 -\nu \dot{\vec u}. \nabla^2 \vec u\nonumber \\
&-& \nu (\nabla^2 \vec u). (\nabla  \phi) + {1\over 2}\nu^2 (\nabla^2 \vec u)^2 + \vec u. \vec J - \nu \vec u . \nabla n -  \phi n,\label{i19}
\end{eqnarray}
and for transforming it from second to first order, we consider the momentum as the auxiliary field, extending the configuration space. In that way, we get the equation of motion for $\vec \pi\,$ : $\vec \pi = - \nabla \phi - \dot{\vec u} + \nu \nabla^2 \vec u$. Replacing back this result in the Lagrangian we return to the original quadratic term. Using that $\vec J = \vec J_{tr} + \vec J_l$, the Lagrangian density in eq.(\ref{i19}) becomes
\begin{eqnarray}
{\cal L} &=& - {1\over 2} \vec \pi^2 - \vec \pi . \nabla \phi - \vec \pi . \dot {\vec u} + \nu \vec \pi . \nabla^2 \vec u + {1\over 2} (\nabla \times \vec u)^2\nonumber \\
&-& \nu \vec u . \nabla n + \vec u . \vec J_{tr} + \vec u . \nabla \dot \phi - n \phi \label{i19a}
\end{eqnarray}

In this case, the first order Lagrangian reads ${\cal L}^{(0)} = - \vec \pi. \dot{\vec u} - \dot{\vec u} . \nabla \phi - U^{(0)}$, where the potential density is $
U^{(0)} = {1\over 2} \vec \pi^2 + \vec \pi . \nabla \phi - {1\over 2} (\nabla \times \vec u)^2 - \nu \vec \pi . \nabla^2 \vec u - \vec u . \vec J_{tr} + \phi n + \nu \vec u . \nabla n$.

The initial set of sympletic variables $\xi_i^{(0)} = \{u_i, \, \pi_i, \, \phi\}\;$ allows us to identify the quantities $
a_i^{(0)\vec u} = - \pi_i - \partial_i \phi,\;
a_i^{(0)\vec \pi} = 0, \hbox{ and }
a_i^{(0)\phi} = 0$.
These lead us to the following matrix $f^{(0)}$, 
\begin{equation}
f^{(0)}= \left( \begin{array}{ccc}
                0 & -\partial_j^y & 0 \\
                \partial_i^y& 0 & \delta_{ij} \\
                0 & -\delta_{ij} & 0 
                \end{array}
         \right) \delta(\vec x - \vec y), \label{i25}
\end{equation}
which obviously is singular. Thus, the system has constraints in the Faddeev-Jackiw formalism. The components of the zero-mode $\tilde v^{(0)} = (v^\phi,\, 0, \, v^{\vec \pi})$ satisfy $v_i^{\vec \pi} - \partial_i v^\phi = 0$. 

The primary constraint is obtained from $\int {\rm d} \vec x v^\phi (\vec x) \left[ - \nabla . \vec \pi (\vec x) + n (\vec x) \right] = 0$. Since $ v^{\phi}(\vec x)\,$ is an arbitrary function and the fluid is taken incompressible,  we obtain the constraint $
\nabla . \vec \pi ( \vec x ) - n(\vec x) = 0 $.

Introducing this constraint back into the Lagrangian by means of a Lagrange multiplier $\lambda\,$ one obtains ${\cal L}^{(1)} = - \vec \pi .\vec u - \nabla \phi . \dot{\vec u} + \dot \lambda (\nabla . \vec \pi - n) - U^{(1)}\,$ where $U^{(1)} = {1\over 2} \vec \pi^2 - {1\over 2}(\nabla \times \vec u)^2 - \nu \vec \pi . \nabla^2 \vec u - \vec u . \vec J_{tr} + \nu \vec u . \nabla n$.

Considering now that the new set of sympletic variables is given in the following order, $\xi = (u_i,\, \phi, \, \pi_i, \, \lambda)$, we have $a_i^{(1)\vec u} = - \pi_i - \partial_i \phi, \; a_i^{(1)\phi} = 0, \, \hbox{ and } a^{(1)\lambda} = \nabla . \vec \pi - n$. From there we obtain the sympletic matrix $f^{(1)}$,  
\begin{equation}
f^{(1)} = \left( \begin{array}{cccc}
                 0 & \partial_i^y & \delta_{ij} & 0 \\
                 - \partial_j^y & 0 & 0 & 0 \\
                 - \delta_{ij} & 0 & 0 & \partial_i^y \\
                 0 & 0 & - \partial_j^x & 0  
                 \end{array}
          \right)\delta(\vec x - \vec y). \label{i33}
\end{equation}
that is a nonsingular matrix. Then we can identify it as the sympletic tensor of the constrained theory. The inverse of $f^{(1)}\,$ gives us the Dirac brackets of the physical fields and can be obtained in a straightforward calculation. The result is 
\begin{equation}
(f^{(1)})^{-1} = \left( \begin{array}{cccc}
                        0 & - {\partial_i^x\over{\nabla^2}}& \delta_{ij} - {\partial_i^x\partial_j^x\over{\nabla^2}} & 0  \\
                        {\partial_j^x\over{\nabla^2}} & 0 & 0 &  {1\over{\nabla^2}} \\
                        - \delta_{ij} + {\partial_i^x\partial_j^x\over{\nabla^2}} & 0 & 0 & {\partial_i^x\over{\nabla^2}}  \\
                        0 & -{1\over{\nabla^2}} &  - {\partial_j^x\over{\nabla^2}} & 0 \\
                        \end{array}
                 \right)\delta(\vec x - \vec y)\label{i42}
\end{equation}

From the matrix $(f^{(1)})^{-1}\,$ we identify
\begin{equation}
\{u_i(\vec x),\, \dot u_j(\vec y)\}^* = \left( -\delta_{ij} + {\partial_i^x\partial_j^x\over{\nabla^2}}\right)\delta(\vec x - \vec y)\label{i43}
\end{equation}
where the auxiliary fields were eliminated and we returned to the original variables. These brackets are nothing more than brackets of the turbulence gauge field in the gauge $\nabla . \vec u = 0$, obtained in eq.~(\ref{i17}), showing us that the Dirac brackets are the same for the Navier-Stokes equation with or without the viscous term.

The straightforward step would be the quantization of this constrained theory, that can be obtained by the well known canonical quantization rule $\{\;, \;\}^* \rightarrow -i [\;, \;]$. Doing so,  we get the commutators
\begin{eqnarray}
\left[ u_i(\vec x), u_j(\vec y)\right] &=& 0 = \left[\pi_i(\vec x), \, \pi_j(\vec y)\right], \nonumber \\
\left[ u_i(\vec x),\pi_j(\vec y)\right] &=& - i \left( \delta_{ij} - {\partial_i^x \partial_j^x\over{\nabla^2}}\right) \delta(\vec x - \vec y).\label{i18}
\end{eqnarray}

Once we have the canonical quantization rule, we can apply standard Quantum Field Theory technics to find the generating functional, consequently correlation functions and all physical quantities \cite{Itzykson,Henneaux} one wish.

\section{Conclusion} 

In conclusion, we have proposed to study incompressible turbulent hydrodynamics in the metafluid dynamics formalism applying tools commonly used in Quantum 
Field Theory for constrained systems. This study is possible because of 
the close analogy between incompressible turbulent hydrodynamics and the electromagnetism. Exploiting this analogy we computed the Dirac brackets using the Dirac and Fadeev-Jackiw formalisms. We showed that Dirac brackets found in the Dirac formalism in the inertial regime ($\nu = 0$) is valid for all regime of fully-developed turbulence, as we found in the Fadeev-Jackiw formalism. We believe this approach of turbulence as a gauge theory can lead to results that tend to renew the  optimism to solve the difficult problem of turbulence.

\section{Acknowledgements} 
This work was partially supported by CNPq and FAPEMIG. One of us (A.C.R.M) thanks to CNPq for the financial support and  (W.O and F.I.T.) would like to thank FAPEMIG and CNPq for partial support.

\end{document}